\documentclass{svjour3}

\smartqed

\usepackage{afterpage}
\usepackage{array}
\usepackage{booktabs}
\usepackage{cite}
\usepackage{enumitem}
\usepackage{float}
\usepackage{listings}
\usepackage{longtable}
\usepackage{multirow}
\usepackage{nicematrix}
\usepackage{graphicx}
\usepackage{tcolorbox}
\usepackage{xcolor}
\usepackage{xspace}
\usepackage{xurl}
\usepackage{hyperref}
\usepackage{todonotes}
\usepackage{lmodern}
\usepackage[normalem]{ulem}
\usepackage{todonotes}
\usepackage{xcolor}
\usepackage[inkscapelatex=false]{svg}
\svgpath{{images/}}
\tcbuselibrary{skins}

\newcolumntype{C}[1]{>{\centering\arraybackslash}p{#1}}

\newcommand{\rpv}[1]{\textcolor{brown}{remove passive voice}}

\definecolor{mikecolor}{RGB}{255,140,0}

\DeclareUnicodeCharacter{202F}{\,}
\extrafloats{500}

\makeatletter
\newcommand{\ie}{\emph{i.e.}\@ifnextchar.{\!\@gobble}{}}
\newcommand{\eg}{\emph{e.g.}\@ifnextchar.{\!\@gobble}{}}
\newcommand{\etc}{\emph{etc}\@ifnextchar.{}{.\@}}
\makeatother

\newcommand{\comm}[1]{}

\newcommand{\rqOne}{RQ1: In what application domains is FMware most commonly used?}
\newcommand{\rqTwo}{RQ2: What technical challenges do developers encounter when building and maintaining FMware?}
\newcommand{\rqThree}{RQ3: Which types of FMware issues require the most time to resolve?}

\newtcolorbox{mybox}[2][]{
top=0.15in,left=4pt,right=4pt,bottom=4pt,
fonttitle=\bfseries,
colbacktitle=gray,
colback=gray!5,
colframe=gray!40!black,
enhanced,
attach boxed title to top left={xshift=0em,yshift=-\tcboxedtitleheight/2},
boxed title style={size=small},
drop shadow={black!50!white},
title=#2,#1}

\journalname{Empirical Software Engineering}

\begin{document}

\titlerunning{An Empirical Study of Developer Challenges and Resolution Times}  
\title{What Slows Down FMware Development?\\ An Empirical Study of Developer Challenges and Resolution Times}

\author{Zitao~Wang\and Zhimin~Zhao \and Michael~W.~Godfrey}


\institute{
    Z. Wang \at 
    School of Computing, University of Waterloo, Waterloo, ON, Canada\\
    \email{z254wang@uwaterloo.ca}
    \and
    Z. Zhao \at
    School of Computing, Queen's University, Kingston, ON, Canada\\
    \email{z.zhao@queensu.ca}
    \and
    M. W. Godfrey \at 
    School of Computing, University of Waterloo, Waterloo, ON, Canada\\
    \email{migod@uwaterloo.ca}
}
\date{Received: date / Accepted: date}

\maketitle

\begin{abstract}
Foundation Models (FMs), such as OpenAI’s GPT, are fundamentally transforming the practice of software engineering by enabling the development of \emph{FMware} --- applications and infrastructures built around these models. 
FMware systems now support tasks such as code generation, natural-language interaction, knowledge integration, and multi-modal content creation, underscoring their disruptive impact on current software engineering workflows. 
However, the design, implementation, and evolution of FMware present significant new challenges, particularly across cloud-based and on-premise platforms where goals, processes, and tools often diverge from those of traditional software development.

To our knowledge, this is the first large-scale analysis of FMware development across both cloud-based platforms and open-source repositories. 
We empirically investigate the FMware ecosystem through three focus areas: (1) the most common application domains of FMware, (2) the key challenges developers encounter, and (3) the types of issues that demand the greatest effort to resolve. 
Our analysis draws on data from GitHub repositories and from leading FMware platforms, including HuggingFace, GPTStore, Ora, and Poe. 
Our findings reveal a strong focus on education, content creation, and business strategy, alongside persistent technical challenges in memory management, dependency handling, and tokenizer configuration. 
On GitHub, bug reports and core functionality issues are the most frequently reported problems, while code review, similarity search, and prompt template design are the most time-consuming to resolve.

By uncovering developer practices and pain points, this study points to opportunities to improve FMware tools, workflows, and community support, and provides actionable insights to help guide the future of FMware development.

\end{abstract}

\section{Introduction}
\label{sec:introduction}
The rapid adoption of Foundation Models (FMs) such as OpenAI's GPT is transforming knowledge-based industries. 
The capabilities of systems built around such models range from personalizing content recommendations in entertainment to enabling predictive diagnostics in healthcare to improving traffic flow in urban planning. 
As a result, foundation models (FMs) are now being embedded into a wide range of software applications, leading to the emergence of \emph{FMware} --- systems and infrastructures built around these powerful models~\cite{Brief}. 
While their potential is considerable, the processes of designing, implementing, and maintaining such systems introduce fundamentally new software engineering challenges, and effective strategies for addressing them remain uncertain.


FMware development involves the organization, deployment, and maintenance of complex models. 
This typically requires managing dynamic environments, processing large volumes of data, and adapting to evolving requirements. 
Both cloud-based and on-premise solutions play important roles in this landscape. 
Cloud-based FMware platforms --- such as GPTstore, HuggingFace, Poe, and Ora --- provide scalable environments that simplify deployment and management, enabling developers to leverage powerful FMs without building systems from the ground up. 
In contrast, on-premise FMware, often hosted on platforms such as GitHub, offers developers greater control and customization, making it easier to tailor models to specialized requirements.

Despite the increasing adoption of FMware in practice, developers continue to face substantial challenges in managing, maintaining, and optimizing these systems. 
Prior studies have documented recurring issues such as prompt instability, hallucinations in generated content, and the high operational costs of deploying large-scale models~\cite{rethinking,tutorialFMware}. 
Further difficulties arise in managing model complexity, ensuring stability, accommodating heterogeneous requirements, and supporting system evolution over time~\cite{tutorialFMware,10.1145/3676151.3719357}.

This study addresses that gap by examining how FMware-related issues emerge across both cloud-based and on-premise environments. 
We focus on identifying the key topics discussed in practitioner communities, characterizing the main challenges developers report, and assessing the time and effort required to resolve issues in FMware repositories. 
Through a systematic analysis of developer discussions and GitHub issue reports, we provide new empirical insights into the FMware ecosystem and highlight opportunities to strengthen tool support, development practices, and community resources.

To develop a deeper empirical understanding of the FMware landscape, we analyzed two complementary data sources: FMware applications deployed on cloud-based platforms (Promptware) and issue reports from GitHub-hosted FMware repositories. 
Specifically, our study investigates the primary application domains of FMware (RQ1), the most significant challenges developers encounter (RQ2), and the types of issues that require the greatest effort to resolve (RQ3):

\begin{itemize}
\item \textbf{\rqOne}
    To answer this question, we analyzed 25,790 FMware descriptions collected from four cloud-based platforms: GPTStore, HuggingFace, Ora, and Poe. 
    Using topic modeling, we extracted recurring themes and organized them into a taxonomy of 33 fine-grained topics grouped under 18 macro-topics.

    Our analysis found that FMware is most often used for education, creativity, and entertainment, with growing traction in professional domains. 
    Specifically, the most prevalent macro-topics are ``Educational Content \& Learning'' (17.7\%), ``Image \& Visual Generation'' (13.8\%), and ``Entertainment \& Gaming'' (12.7\%); ``Content Creation \& Writing'' (11.8\%) and ``Business \& Strategy'' (10.8\%) further highlight adoption in professional contexts.

\item \textbf{\rqTwo}
    We analyzed 138,989 GitHub issues from 669 FMware repositories and automatically classified them into 15 categories. 
    To gain deeper insight, we applied topic modeling on the two largest categories --- ``Bug Reports'' and ``Core Functionality'' --- to identify common pain points in development.

    Our analysis shows that FMware development remains technically complex and error-prone. 
    The most frequently reported issues were ``Bug Reports'' (30.7\%) and ``Core Functionality'' (21.9\%), with recurring challenges in memory management, dependency conflicts, tokenizer context length, and model loading failures. 
    Together, these findings point to a pressing need for better tooling and debugging support.

\item \textbf{\rqThree}
    We examined over 30,000 resolved issues across the ``Bug Reports'' and ``Core Functionality'' categories and measured their resolution times, aggregating results by topic to identify which issues demand the greatest developer effort.

    Our analysis found that ``Code Contribution and Review Processes'' (934.2 hours), ``Similarity Search and Vector Stores'' (787.5 hours), and ``Prompt Templates for Model Interactions'' (654.8 hours) required the most time to resolve. 
    These findings highlight collaborative workflows, retrieval-augmented architectures, and user-facing prompt design as major bottlenecks in FMware engineering.
\end{itemize}


\section{Background and Related Work}
\label{sec:background}

In this section, we provide an overview of FMware. 
Conceptually, FMware can be divided into two categories: \textit{Promptware}, which relies on natural-language prompts to generate outputs, and \textit{Agentware}, which involves autonomous agents capable of decision making and task execution. This distinction highlights how FMware departs from traditional software development practices, where code-based logic designed by humans was the primary method of building systems.  

For the purpose of this study, however, we adopt a deployment-oriented view of FMware: (i) cloud-based \textbf{Promptware platforms} (e.g., GPT Store, Hugging Face, Poe, Ora), and (ii) \textbf{GitHub-hosted FMware} in open repositories. These two environments represent the primary ways FMware is currently developed and maintained, and they form the focus of our empirical analysis.  

We next examine the unique challenges posed by FMware, including the complexities of prompt engineering, model alignment, and managing system behaviors. Finally, we narrow our focus to the resolution times of issues observed in GitHub repositories. This framing sets the stage for our empirical analysis in the following sections.

\subsection{FMware Introduction and Classification}

Over the past decade, software engineering has undergone a series of transformative shifts~\cite{cico2021exploring}. Traditional software systems --- often termed \emph{Codeware} --- are built from manually written logic expressed in conventional programming languages. The incorporation of machine learning into development workflows introduced \emph{Neuralware}, in which data and AI models are treated as first-class development artifacts~\cite{hassan2024ainativesoftwareengineeringse,8804457}. Most recently, the rise of systems centered on Foundation Models (FMs) has ushered in the age of \emph{FMware}: a broad class of software that harnesses the generative and interactive capabilities of large pre-trained models~\cite{rethinking,tutorialFMware}.  

Conceptually, FMware can be divided into two categories: \emph{Promptware}, which enables user interaction through natural-language prompts, and \emph{Agentware}, which employs autonomous agents capable of multi-step reasoning and decision-making~\cite{liu2023pre,wang2024survey}. While this distinction captures the diversity and ongoing evolution of FMware, our study focuses on a deployment-oriented perspective: (i) cloud-based Promptware platforms and (ii) GitHub-hosted FMware repositories.  

Despite their promise, FMware systems introduce new engineering challenges, such as hallucinations, unrepeatable outcomes, constraints when interacting with external resources, and the complexity of multi-step task orchestration~\cite{Brief, rethinking,fan2023largelanguagemodelssoftware}. Addressing these issues requires a rethinking of the software lifecycle --- from requirements and design through implementation, coordination, and operations --- to accommodate the cognitive and architectural demands introduced by FMs~\cite{langchain2024stateofagents,raji2021ai,yao2023tree}.

\subsection{Challenges in FMware Development}

\subsubsection{Technical Challenges} 

Prompt instability and hallucinations --- where FMs confidently produce incorrect output --- are significant technical challenges affecting the development of FMware~\cite{Hallucination,chen2025promptwareengineeringsoftwareengineering,chen2025unleashing,subramonyam2025prototypingpromptsemergingapproaches}. These behaviours introduce unpredictability and, for complex multi-step tasks, necessitate careful prompt engineering and orchestration~\cite{yao2023tree}. In cloud-based FMware platforms such as GPTstore and HuggingFace, developers must contend with unpredictable model behaviours while having limited control over model versions and inference settings. By contrast, teams building on hosted, open-source FMware projects --- typically deployed via GitHub --- gain greater customization and low-level control of model configurations and dependencies, but assume additional operational burdens for configuration, evaluation, maintenance, and monitoring~\cite{rethinking,tutorialFMware}.

\subsubsection{Operational Challenges}

Operationally, managing large-scale models entails substantial resource and cost pressures, especially when aligning them with enterprise-specific data.\cite{patterson2021carbon} These pressures are compounded by the need to ensure scalability and stability in cloud environments while preserving flexibility and control in on-premise FMware. Practitioners also struggle to monitor and assure performance and reliability, as the stochastic nature of FMs hinders reproducibility, complicates incident triage, and slows troubleshooting during operations and maintenance~\cite{rethinking,gundersen2018state}. Understanding these operational constraints is essential to strengthening the FMware ecosystem, as reflected in issues frequently reported by developers on platforms such as GitHub.

\subsection{State of FMware Tools and Platforms}

Building on the deployment-oriented classification introduced earlier, we focus on two primary environments where FMware is currently developed and maintained: \textbf{Promptware platforms} and \textbf{GitHub-hosted FMware}. These categories reflect the settings analyzed in our empirical study.  

\subsubsection{Promptware Platforms}

Promptware platforms, such as GPTStore and Hugging Face Spaces, provide plug-and-play environments where users can interact with pre-trained FMs through natural-language prompts. By lowering technical barriers and supporting rapid customization, these platforms have enabled a wide range of applications, from education to business automation~\cite{shen2024hugginggpt,gptstore,jiang2025prompt}. However, they also introduce challenges, including scalability constraints, high computational costs, and reliance on third-party APIs.  

\subsubsection{GitHub-hosted FMware}

In contrast, GitHub-hosted FMware projects provide developers with direct access to FM implementations, offering greater customization, fine-tuning, and control over model behaviors. These repositories often include configurations, API integrations, and deployment frameworks, but also require teams to manage dependencies, optimize large models, and operate resource-intensive infrastructure~\cite{rajbahadur2024cool,alhanahnah2024depsragagenticreasoningplanning,reyes2025byamfixingbreakingdependency}. While this approach affords flexibility and enterprise-level control, it introduces significant operational and maintenance burdens.  

\subsubsection{Motivation}

Despite the rapid growth of both types of FMware, a significant gap remains in understanding the practical challenges developers face when working in these environments. Previous research has largely discussed conceptual or theoretical concerns (e.g., prompt instability, hallucination)~\cite{rethinking,survey1,survey2}, but less is known about the day-to-day issues encountered in real-world Promptware platforms and GitHub-hosted repositories. Our study seeks to bridge this gap by conducting an empirical analysis of prevalent topics, developer challenges, and issue resolution times across these two environments.

\section{Methodology}
\label{sec:methodology}

This section introduces our research goals and questions, and provides a detailed explanation of our study design.
\subsection{Goal and Research Questions}
\label{sec:method:goal}


Our study seeks to characterize the principal concerns and technical challenges in FMware development and maintenance. As FMware assumes a growing and critical role in software solutions, it becomes increasingly important to be able to identify the topics that are most relevant to practitioners and the obstacles they encounter across the development lifecycle.  We examine two platform scenarios: (i) cloud-based Promptware platforms (e.g., GPT Store, Hugging Face, Poe, Ora) and (ii) GitHub-hosted FMware in public repositories.  We selected these types of platforms because of their popularity and the size of their active communities. By analyzing both types of environment, we aim to provide insights that generalize across the FMware landscape and support more effective development and maintenance practices.

To address this, we developed three research questions to guide our study:

\begin{itemize}

\item\textbf{RQ1: In what application domains is Promptware most commonly used?}

We aim to identify the major application domains discussed within Promptware platforms. 
By applying topic modeling to extracted app descriptions, we can examine which domains are most prominent across different platforms. Addressing this question provides insight into the areas that developers are currently engaging with and highlights opportunities for improvement and innovation. Understanding these prevalent application domains can also help practitioners and researchers better align their efforts with the active needs and interests of the FMware community.

\item\textbf{RQ2: What technical challenges do developers encounter when building and maintaining FMware?}

We aim to improve understanding of the specific challenges and difficulties that developers encounter when working with FMware. By analyzing issues from GitHub, we aim to uncover the real-world problems that practitioners face during development and maintenance. This is important because while RQ1 provides an overview of popular topics, RQ2 delves into the practical difficulties that developers experience, offering insights into the barriers that need to be addressed to improve the FMware ecosystem.

\item\textbf{RQ3: Which types of FMware issues require the most time to resolve?}

We examine the median and mean resolution times for the issues identified in RQ2. 
By analyzing which kinds of issues take the most time to resolve, we aim to highlight the topics that demand the most effort and resources from developers. 
This understanding can inform the FMware community about where support, documentation, or tooling improvements are most needed, helping to reduce the overall maintenance burden on practitioners.

\end{itemize}

Overall, our study aims to map the application domains of FMware (RQ1) and to uncover the practical and time-intensive challenges of FMware development (RQ2, RQ3), providing insights to guide practitioners, researchers, and tool builders in strengthening the ecosystem.


\subsection{Study Design}
\label{sec:method:design}

\begin{figure}[!t]
\centering
\includegraphics[width=\columnwidth]{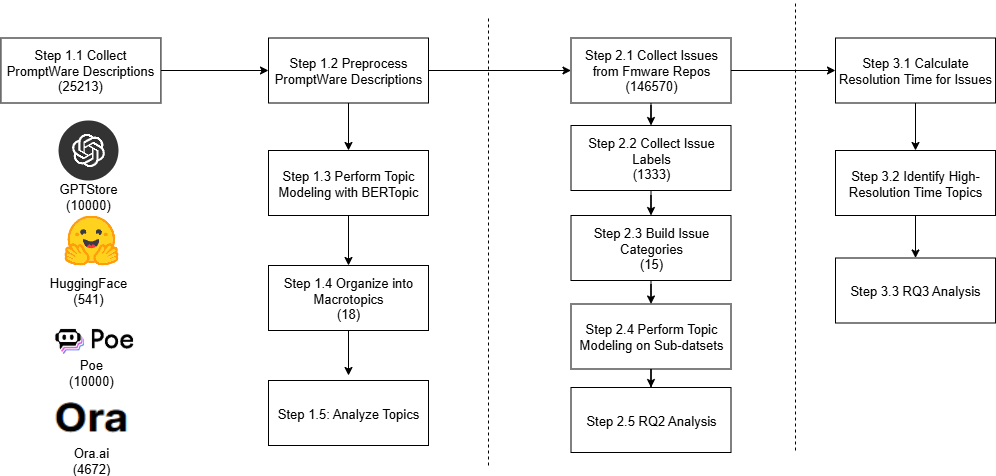}
\vspace{-2em}
\caption{Study workflow to analyze the Promptware descriptions and FMware challenges.}
\label{fig:study workflow}
\vspace{-1.5em}
\end{figure}

We present the workflow of our study in Figure \ref{fig:study workflow}. 
To investigate RQ1, we begin with the collection of cloud-based FMware descriptions from popular closed-source platforms.
Next, we preprocess these collections of descriptions and use the BERTopic technique to perform topic extraction~\cite{grootendorst2022bertopic}. 
We then group the results into ``macrotopics'' and compare the different distributions across different platforms. 
To investogate RQ2, we then collect issues from the FMware community on GitHub. 
Based on the issue tags given by the developers, we build a series of categories for the issues and analyze the distribution of different types of issues. Then we perform topic modeling on those categories in which developers and users may have interests. 
Finally, to investigate RQ3 we evaluate and compare the median and mean solving time for issues and clarify what topics of issues require the most developer efforts to resolve.
\subsubsection{Phase 1: Topic modeling for PromptWare descriptions}
\label{sec:method:design:phase1}

In this phase, our goal is to collect a set of FMware systems and their descriptions from multiple cloud-based close-source platforms. We select our dataset based on the population of PromptWares and popularity of the platforms. From this data collection, we applied topic modeling which yielded the distribution and ``hot'' trends in closed-source FMware.\\

\noindent \textbf{Step 1.1: Collect PromptWare descriptions}\\[-0.5em]

We begin by reviewing a range of popular closed-source FMware platforms in the market such as GPTStore. Our goal is to identify FMware platforms that host general-purpose FMware systems, similar to how traditional app stores offer a wide range of applications. After filtering, we chose three top platforms based on the number of available FMware applications listed (GPTStore, POE, and Ora), and one platform based on its broad popularity and recent rapid growth (the HuggingFace Assistant platform) as the basis of our dataset.
In the end, we collected 90,383 app descriptions from GPTstore, 12,380 from Poe, 4,672 from Ora, and 541 from HuggingFace.

We select the data set for topic modeling to ensure a balance between representativeness and computational feasibility. The data set includes 10,000 app descriptions from both GPTstore and Poe, together with the full datasets from both Ora (4,672 app descriptions) and HuggingFace (541 app descriptions).
The decision to select 10,000 apps from the two larger platforms provides a robust yet manageable sample size for capturing the diversity of apps without overwhelming our computational resources. Meanwhile, the smaller datasets from Ora and HuggingFace were included in full due to their relatively small size but likely significant value, particularly HuggingFace, which is rapidly expanding in both app count and popularity.  
This selection ensures that we have sufficient data for meaningful topic analysis across different FMware platforms while maintaining a practical scope for the analysis.\\[-0.75em]

\noindent \textbf{Step 1.2: Preprocess PromptWare descriptions}\\[-0.75em]

After extracting PromptWare descriptions from the selected platforms, we preprocess the text using standard natural language processing (NLP) techniques:

\begin{enumerate}
    \item Tokenize the input stream. 
    \item ``Scrub'' the tokens to eliminate non-alphabetic characters such as numbers and punctuation marks.
    \item Using the \texttt{spaCy} NLP library, identify and retain only \textit{nouns} and \textit{verbs}, as they are the most meaningful parts of speech for our analysis~\cite{spacy}.
    \item Eliminate unnecessary words to enhance topic modeling clarity. We categorize stop words into three distinct types:
        \begin{enumerate}
            \item \textbf{General stop words:} Common words in English (e.g.,``you'', ``like'', ``the'') that do not contribute to meaningful topic extraction.
            \item \textbf{Platform-specific words:} Words that occur frequently in the FMware descriptions but contribute little discriminatory value, such as “GPT”, “GPTStore”, and “POE”.  
            We extracted the most frequent words from FMware descriptions and manually reviewed them to filter out terms that were highly repetitive yet semantically uninformative for topic modeling. Words that were excessively platform-branded (e.g., “GPTStore”) or redundant in distinguishing meaningful topics were added to the stop word list.
            \item \textbf{FMware-specific terms:} Words closely associated with FMware functionalities but not beneficial for topic modeling, such as “AI”, “Bot”, and “Chatbot”.
        \end{enumerate}
        We evaluate each stop word by analyzing coherence scores and reviewing the top 10 keywords for each topic. A full list of our stop words can be found in the replication package.\footnote{\url{https://github.com/Nick77666/FMware-replication-package}}
\end{enumerate}

Following these steps, the preprocessed text is ready for input into BERTopic for hyperparameter tuning and subsequent topic extraction.\\[-0.75em]

\noindent \textbf{Step 1.3: Perform Topic Modeling with BERTopic}\\[-0.75em]

We employ BERTopic\footnote{\url{https://github.com/MaartenGr/BERTopic}} to perform topic modeling on the preprocessed PromptWare descriptions. BERTopic offers several advantages over traditional topic modeling approaches, making it particularly well-suited for our analysis. For example, unlike models such as bag-of-words or n-grams, which are sparse and high-dimensional, BERTopic leverages \textit{BERT embeddings} to capture rich semantic nuances in text.

BERT (Bidirectional Encoder Representations from Transformers) embeddings are dense vector representations of text generated by a transformer-based model. Unlike traditional embeddings, BERT captures both left and right word context, improving semantic understanding. This approach has been found to enhance topic coherence, leading to more meaningful topic clusters~\cite{grootendorst2022bertopic}.

To optimize the performance of BERTopic, we conduct comprehensive hyperparameter tuning. Guided by the  BERTopic documentation as well as insights from its GitHub community, we experiment with parameters such as \textit{n\_grams}, \textit{min\_cluster\_size}, and \textit{n\_components}. Additionally, we incorporate the top five embedding models from HuggingFace's Massive Text Embedding Benchmark (MTEB) leaderboard\footnote{\url{https://huggingface.co/spaces/mteb/leaderboard}} to identify the best-suited embeddings for our dataset.

During the tuning process, we evaluate the generated topics using Coherence C\_V and Coherence UMass scores~\cite{syed2017full}, which are widely-adopted metrics for assessing the quality of topic models. If coherence scores improve significantly at the boundaries of the hyperparameter search space, we heuristically expand the search range for further refinement. Conversely, we halt adjustments when performance begins to degrade, ensuring stability and reliability in the final model.

In cases where coherence scores are similar across different configurations, we consider additional metrics, such as the number of topics generated and the proportion of descriptions that align with identified topics. To further ensure topic quality, we manually review a random sample comprising 5\% of the clustered descriptions from each topic. This step helps validate the semantic relevance and consistency of the topics.

If topics appear unclear, overly broad, or exhibit significant overlap, we revisit earlier preprocessing steps or refine hyperparameter settings. This iterative approach ensures that the final topics are both distinct and interpretable, providing actionable insights into the thematic structure of PromptWare descriptions. By integrating robust tuning practices and manual validation, we enhance the reliability and utility of our topic modeling results.\\[-0.75em]
 
\noindent \textbf{Step 1.4: Organize into Macrotopics} \\[-0.75em]

After applying BERTopic, we identified a total of 33 distinct topics from the FMware descriptions. To enhance interpretability and reduce redundancy, we aggregated these topics into 18 macrotopics. This process involved grouping semantically similar topics that shared common themes or closely related subject matter. We used the ``negotiated agreement'' technique~\cite{campbell2013coding} to enhance the reliability of the coding process.  This approach involved collaborative coding among the authors, ensuring consensus and refining topic definitions; similar techniques have been effectively employed in prior studies, such as those by Chen et al.~\cite{chen2023practitioners}, demonstrating its value in improving coding accuracy and consistency. 

By consolidating overlapping or narrowly defined topics, the resulting macro-topics provide a higher-level and more structured representation of the dataset, making it easier to identify dominant themes and trends. This aggregation reduces fragmentation, ensures meaningful topic differentiation, and allows for a clearer analysis of the FMware ecosystem.\\[-0.75em]

\noindent \textbf{Step 1.5: Analyze Topics}\\[-0.75em]

After consolidating the BERTopic output into 18 macrotopics, we analyzed the results to determine the most common application domains of Promptware. Specifically, we examined the distribution of Promptware descriptions across the macrotopics to identify which domains were most prevalent. To verify topic quality, we reviewed representative descriptions within each macrotopic to check that they were semantically consistent with the assigned category. This analysis provided a clear mapping from raw topic clusters to interpretable application domains, allowing us to report dominant areas of Promptware usage and establish a baseline for further comparison with FMware repositories in Phase 2.




\subsubsection{Phase 2: Topic modeling for Github Issues from FMware repositories}
\label{sec:method:design:phase2}
In this phase, our focus shifts to gathering GitHub issues from FMware repositories to further analyze the challenges, feature requests, and overall developer experiences related to FMware. By examining these issues, we aim to gain deeper insights into the practical concerns and improvements raised by users and contributors. These insights complement the trends identified in Phase 1 by highlighting the real-world problems developers face when using FMware and how these issues align with the broader topics identified through topic modeling. This phase involves systematically collecting and categorizing issues from selected GitHub repositories to ensure comprehensive coverage of the common challenges and opportunities for enhancement within FMware ecosystems.\\[-0.75em]

\noindent \textbf{Step 2.1: Collect FMware Repositories}\\[-0.75em]

The primary motivation for collecting FMware repositories is to establish a baseline for comparing FMware (from GitHub repositories) with Promptware applications (from platforms such as GPTStore, HuggingFace, POE, and Ora). FMware systems often emphasize technical depth, open-source collaboration, and community-driven development, while Promptware systems focus on user-facing applications and general-purpose functionalities. By analyzing both, we aim to highlight the similarities and differences between these two ecosystems and reinforce the validity of our topic-modeling results. Additionally, GitHub repositories provide rich metadata, such as README files and issue discussions, which offer deeper insights into developer challenges and priorities.

To collect FMware repositories, we combined two approaches. First, we systematically searched GitHub using domain-specific tags and keywords, such as GPT,'' LLM,'' FMware,'' and AI tools.'' This approach ensured a wide coverage of repositories relevant to FMware development. Second, we leveraged GitHub's ``awesome lists'' to identify curated collections of FMware projects. These lists are a valuable resource for discovering high-quality repositories across various domains. A full list of the search terms is provided in the Appendix.

After collecting an initial candidate set of 2,622 repositories, we applied rigorous exclusion criteria to refine the dataset:
\begin{itemize}
\item \textbf{Inactive Repositories:} Excluded repositories not updated in the last 12 months.
\item \textbf{Lack of Documentation:} Excluded repositories without at least minimal documentation (e.g., a README file or contributor guide).
\item \textbf{Non-English Repositories:} Excluded repositories primarily documented in languages other than English.
\item \textbf{Non-FMware Repositories:} Excluded repositories that, after manual inspection, were deemed unrelated to FMware development.
\item \textbf{Low Engagement Repositories:} Excluded repositories with zero stars or forks, indicating little community interaction or activity.
\end{itemize}

Applying these filters reduced the dataset from 2,622 candidates to a final set of 669 valid repositories. These FMware repositories serve as a baseline for comparing their functionality and themes with Promptware descriptions obtained earlier in Step 1.1. This baseline enhances the robustness of our topic modeling approach, allowing us to validate whether the topics generated from Promptware datasets align with or diverge from those observed in the FMware domain.

\noindent \textbf{Step 2.2: Collect Issues}\\[-0.75em]

In this step, we focus on collecting issues from the FMware repositories identified in Step 2.1. GitHub issues provide information on challenges, bugs, feature requests, and user feedback related to FMware development. We use the following approach for issue collection.
\begin{itemize}
    \item \textit{Filtering active repositories}: We target repositories that meet the inclusion criteria described in Step 2.1.
    \item \textit{Issue retrieval}: For each repository, we collect both open and closed issues. Closed issues represent resolved challenges or completed feature requests, while open issues highlight ongoing problems or discussions.
    \item \textit{Issue metadata}: Along with the issue content, we gather metadata such as creation date, labels, number of comments, participants, open time, and closing time. By analyzing the open and close times, we can measure how long it takes to resolve each issue, providing insights into the efficiency and responsiveness of the development process.
\end{itemize}

\noindent We concluded this step with a total of 146,570 valid issues collected from the FMware repositories. These issues, along with their associated metadata and resolution times, form the foundation for our analysis of the challenges, trends, and responsiveness within FMware development.\\

\noindent \textbf{Step 2.3: Open Card Sorting to Build Issue Categories}

In this step, we applied the open card sorting method to organize the 1,333 unique issue labels extracted from the dataset in Step 2.2.\cite{zimmermann2016card} These labels, which span bug reports, feature requests, documentation tasks, and more, represent a diverse set of developer concerns. To conduct the card sorting, the first two authors independently reviewed and grouped the issue labels into thematic clusters based on semantic similarity.

After the initial independent groupings, we conducted a reconciliation session using the negotiated agreement technique. During this session, the authors compared their categorizations, discussed disagreements, and refined the groupings through consensus. We calculated the initial inter-coder agreement using Cohen’s Kappa, which yielded a moderate agreement score of k = 0.71, indicating reasonably high consistency between the two reviewers. Disagreements were resolved through discussion until full agreement was reached on all labels.

Through this iterative process, we finalized a taxonomy of 15 distinct issue categories that best represent the themes emerging from the FMware issue dataset. This structured classification allows us to focus our analysis on key areas relevant to FMware development, such as bug reports, core functionality issues, and performance optimization. The resulting taxonomy serves as a foundational layer for downstream topic modeling and resolution time analysis in subsequent steps of our study.\\

\noindent \textbf{Step 2.4: Mapping Issues to Categories Using Labels and AI-Assisted Classification}\\[-0.75em]

In this step, we map each issue to one of the 15 predefined categories that we found in the previous step, leveraging issue labels where available. For the \textrm{48,700 labeled issues}, we directly map them to the appropriate category based on their labels. However, \textrm{98,870 issues} in our dataset were unlabelled, necessitating an additional step for classification.

To categorize these unlabeled issues, we utilized the \textbf{Claude 3-Haiku model}. This AI model was used to classify the issues based on both the issue title and issue body. The model was prompted with a predefined set of 15 categories, and for each issue, it classified the content by matching it to one of these categories based on semantic analysis. We use the following prompt for the issue classification:
\begin{tcolorbox}[colback=gray!5!white, colframe=gray, title=Classification Instruction]
\begin{verbatim}
You are a professional SE researcher and categorize issues 
accurately. Classify the following Github issue titles 
into these topics by number:
1. Bug Reports
2. Feature Requests
3. Documentation
4. User Experience (UX)
5. Security
6. Core Functionality
7. Project Management
8. Quality Assurance & Testing
9. Infrastructure & Deployment
10. Support & Questions
11. API & Integration
12. Platform Compatibility
13. Performance & Optimization
14. Miscellaneous
15. Internationalization & Localization

For each issue, respond with the issue number 
and the corresponding topic number, one per line.

Titles and Bodies:
\end{verbatim}
\end{tcolorbox}

To assess the reliability of AI-assisted classification, we manually validated 200 randomly sampled issues. Agreement between AI labels and human judgment was 86\%, suggesting that the approach is reasonably reliable for large-scale categorization, though some misclassification remains possible.

This method ensured that even the issues without pre-existing labels were systematically categorized, allowing us to maintain a comprehensive and consistent dataset for further analysis. The combination of label-based and AI-assisted categorization enabled us to handle the large volume of data effectively.\\[-0.75em]

\noindent \textbf{Step 2.5: Preprocessing Issues for Topic Modeling}\\[-0.75em]

To prepare the Bug Reports and Core Functionality issues for topic modeling, we applied a similar preprocessing pipeline as described in Step 1.2 for PromptWare descriptions. We reused the same NLP framework and processing flow, with slight adjustments (indicated by \emph{italicized text}) tailored to GitHub issue data:

\begin{enumerate}
    \item Tokenize the input stream.
    \item ``Scrub'' the tokens to eliminate non-alphabetic characters such as numbers and punctuation marks.
    \item Using the \texttt{spaCy} NLP library, identify and retain only \textit{nouns} and \textit{verbs}, as they are the most meaningful parts of speech for our analysis~\cite{spacy}.
    \item Eliminate unnecessary words to enhance topic modeling clarity. We categorize stop words into three distinct types:
        \begin{enumerate}
            \item \textbf{General stop words:} Common words in English (e.g., ``you'', ``like'', ``the'') that do not contribute to meaningful topic extraction.
            \item \textbf{Discussion-specific words:} \textit{Words such as ``discuss'', ``post'', and ``difference'' that occur frequently in GitHub issue discussions but add little value for identifying topics.}
            \item \textbf{Preserved diagnostic terms:} \textit{Unlike the PromptWare descriptions, we retained FMware-relevant terms such as ``error'', ``bug'', and ``crash'' because they carry essential diagnostic information.}
        \end{enumerate}
        \textit{We refined the stop word list by evaluating topic coherence scores and manually inspecting top keywords, as in Step 1.2.}
\end{enumerate}

As with PromptWare descriptions, we refined the stop word list through coherence score evaluation and manual review of top keywords per topic. The final preprocessed text was then input into BERTopic for modeling. For reproducibility,the complete stop word list is provided in the Appendix.\ref{app:Rq2stopwords}.\\

\noindent \textbf{Step 2.6: Apply Topic Modeling to Selected Categories}\\[-0.75em]

After preprocessing the issues in Step 2.5, we applied BERTopic to perform topic modeling on the two selected categories: Bug Reports and Core Functionality. Following the approach outlined in Step 1.5, we identified distinct topics and then aggregated them into broader macro-topics to reduce redundancy and improve interpretability. This process allowed us to group semantically similar topics, ensuring that the final set of topics was comprehensive and distinct.

We then evaluated these macro-topics based on their ``prevalence,'' consistent with the method described in Step 1.6. Prevalence, which measures the frequency with which a topic appears in the dataset, helped us prioritize the most relevant macro-topics for deeper analysis. In addition, we identified low prevalence topics that may still hold significance in niche contexts. This evaluation provided a clearer understanding of the key themes within Bug Reports and Core Functionality, guiding the next steps of our qualitative and quantitative assessments.\\

\subsubsection{Phase 3: Analyzing Issue Resolution Time by Topic}
\label{sec:method:design:phase3}

In this phase, we address RQ3: Which topics of issues require more solving time for developers? Based on the topics identified in RQ2, we analyze the median and mean resolution time for each topic to evaluate which categories demand the most developer effort.

We begin by calculating the resolution time for each issue, defined as the time between the issue's opening and closing. This data is then grouped by the topics generated through topic modeling in Phase 2. Our focus is on categories that developers and users are likely to prioritize, such as Bug Reports and Core Functionality, as these tend to attract the most attention and have a significant impact on FMware development.

By analyzing the median and mean solving time across different topics, we aim to identify which issues require the greatest investment of time and resources from developers. This analysis will help highlight the areas where developer efforts are most concentrated and provide insights into potential bottlenecks in issue resolution. Understanding which topics take the longest to resolve will inform strategies for optimizing developer workflows and improving the overall efficiency of FMware development.\\[-0.75em]

\noindent \textbf{Step 3.1: Calculate Resolution Time for Each Topic}\\[-0.75em]

In this step, we calculate median and mean resolution time for each topic. For each issue, the resolution time is determined by taking the difference between the issue's open time and close time. This time difference reflects how long it took for the issue to be resolved. Once we have the resolution time for all issues, we group them by their assigned topics from Phase 2.

For each topic, we then compute the median and mean resolution time to identify how long issues in that category typically take to resolve. 
We primarily focus on the median as it is more robust to outliers and better represents the typical resolution time. 
The mean is also reported to provide additional context where relevant.\\[-0.75em]

This step will provide a clear understanding of which topics require more or less time for developers to address, helping us to identify patterns and potential bottlenecks in the development process.\\[-0.75em]

\noindent \textbf{Step 3.2: Identify High-Resolution Time Topics}\\[-0.75em]

Once the resolution time for all topics is calculated, we proceed by ranking the topics based on their median resolution time. We also examine the topics with the highest mean resolution times for comparison, offering supplementary insight into topics with highly variable or prolonged resolution durations. These high-resolution time topics likely represent complex or resource-intensive issues that may warrant further attention or process improvements.

This ranking allows us to identify which topics are more time-consuming for developers to resolve.
By highlighting the topics with the highest mean resolution times, we can pinpoint areas where developers are spending the most effort and time. These 
high-resolution time 
topics likely represent complex or resource-intensive issues that may require further investigation or prioritization for process improvements. 
This step thus helps to focus attention on potential bottlenecks or areas that may need additional resources to reduce issue resolution times.

\section{Results}
In this section, we present the results of our investigation into our three research questions.

\subsection{RQ1 Results} 
\label{sec:rq1}

\textbf{RQ1: In what application domains is Promptware most commonly used?}  

To answer this question, we examine the distribution of macro-topics across various FMware platforms to understand the differing focuses and interests within the closed-source FMware ecosystem. We begin by categorizing FMware descriptions into 33 fine-grained topics, which are then aggregated into 18 macro-topics through manual labeling based on semantic similarity and domain knowledge. This classification helps to reveal high-level trends across the ecosystem.

Each macro-topic is accompanied by a clear definition and an illustrative example to highlight its main characteristics. All examples are direct quotes from the FMware descriptions in our dataset, selected to best represent the theme of each topic. Table~\ref{tab:macro_topic_prevalence_ids} presents the relative frequency of each macro-topic across platforms. A complete list of the 33 fine-grained topics, their IDs, and representative examples is included in the Appendix for reference.


\begin{table}[ht]
\centering
\begin{tabular}{llr}
\toprule
\textbf{Macro-topic} & \textbf{Topic IDs} & \textbf{Prevalence (\%)} \\ 
\midrule
Educational Content \& Learning & 1, 6, 28, 30 & 17.74 \\ 
Image \& Visual Generation & 0 & 13.78 \\ 
Entertainment \& Gaming & 3, 12, 16, 19, 29 & 12.69 \\ 
Content Creation \& Writing & 7, 8, 9, 23, 24 & 11.78 \\
Business \& Strategy & 2, 22, 33 & 10.75 \\ 
Technology \& Programming & 4 & 6.11 \\ 
Health \& Wellness & 5 & 5.62 \\ 
Language \& Translation & 11 & 2.33 \\ 
Travel \& Tourism & 13 & 2.07 \\ 
Coaching \& Advice & 14 & 1.93 \\ 
Chatbots \& Interaction & 15 & 1.88 \\ 
Communication \& Dialogue & 17 & 1.49 \\ 
Science \& Exploration & 18 & 1.49 \\ 
Art \& Music & 20 & 1.47 \\ 
Repair \& Maintenance & 21 & 1.42 \\ 
Agriculture \& Farming & 26 & 1.08 \\ 
Culture \& History & 31 & 0.88 \\ 
Legal \& Documentation & 32 & 0.88 \\ 
\bottomrule
\end{tabular}
\caption{Prevalence of macro-topics and their associated topic IDs in the dataset}
\label{tab:macro_topic_prevalence_ids}
\end{table}

\mbox{}\\
\noindent \textbf{Definitions of Macro-topics:}

\begin{itemize}
\item \textbf{Educational Content \& Learning} --- Topics related to educational tools, learning resources, and platforms designed to enhance knowledge and skills in various domains.
    \begin{description}
    \item[\textbf{Example:}] \textit{ Your personal AI tutor by Khan Academy! I'm Khanmigo Lite - here to help you with math, science, and humanities questions. I won’t do your work for you, but I will help you learn how to solve them on your own. Can you tell me the problem or exercise you’d like to solve?}
    \end{description}
\item \textbf{Image \& Visual Generation} --- Topics that focus on tools and applications for generating, editing, and enhancing images and visuals, often using AI or machine learning techniques.
    \begin{description}
    \item[\textbf{Example:}] \textit{ A GPT specialized in generating and refining images with a mix of professional and friendly 
            tone.image generator}
    \end{description}

\item \textbf{Content Creation \& Writing} --- Topics that encompass tools
    and platforms for writing, content generation, and creative expression,
    including blogs, articles, and storytelling.
    \begin{description}
    \item[\textbf{Example:}] \textit{ A creative AI assistant for content
	creation and writing.}
    \end{description}

\item \textbf{Business \& Strategy} --- Topics that include applications
    and tools for business management, strategic planning, and professional
    development.
    \begin{description}
    \item[\textbf{Example:}] \textit{Get tailored strategies for your
	business growth and online success! Tackle everything from business
	plans to digital marketing trends! } 
    \end{description}

\item \textbf{Technology \& Programming} --- Topics that are centered on
    programming, software development, and technological advancements,
    including coding practices, software tools, and innovative tech
    solutions. 

    \begin{description}
    \item[\textbf{Example:}] \textit{Delivers complete solutions to any
	programming problem, no matter the language and the complexity.}
    \end{description}

\item \textbf{Health \& Wellness} --- Topics that relate to health,
    fitness, wellness, and healthcare tools designed to promote physical
    and mental well-being.

    \begin{description}
    \item[\textbf{Example:}] \textit{Create your own healthy recipes with
	nutritional insights, it can help you create a balance diet plan.}
    \end{description}

\item \textbf{Language \& Translation} --- Topics that focus on language
    processing, translation services, and multilingual communication tools
    that facilitate cross-language interaction. 

    \begin{description}
    \item[\textbf{Example:}] \textit{ Accurate and natural like human
	translations in all major languages}
    \end{description}

\item \textbf{Travel \& Tourism} --- Topics that cover tools and resources
    related to travel planning, tourism, and exploration, including travel
    guides, booking systems, and travel recommendations.

    \begin{description}
    \item[\textbf{Example:}] \textit{ Expert on global travel destinations,
	trip planning, budget building, and exploring the world! Press T
	for Travel Menu.}
    \end{description}

\item \textbf{Coaching \& Advice} --- Topics that include personal
    development, coaching, and advisory services, offering guidance on
    various aspects of life and career.

    \begin{description}
    \item[\textbf{Example:}] \textit{ Smart Career Coach for career advice,
	career tips, career coaching.}
    \end{description}

\item \textbf{Chatbots \& Interaction} --- Topics that involve
    conversational agents, chatbots, and interactive dialogue systems
    designed to enhance user engagement and automate communication.

    \begin{description}
    \item[\textbf{Example:}] \textit{ A confident and witty chatbot for
	fun, respectful romance chats.} 
    \end{description}

\item \textbf{Communication \& Dialogue} --- Topics that encompass tools
    for enhancing communication, dialogue, and social interaction,
    facilitating effective conversation and information exchange.

    \begin{description}
    \item[\textbf{Example:}] \textit{ Creates one-question, one-answer
	animal dialogue comics.}
    \end{description}

\item \textbf{Science \& Exploration} --- Topics that are Related to
    scientific inquiry, research, and exploration across various fields of
    science, promoting discovery and knowledge expansion.

    \begin{description}
    \item[\textbf{Example:}] \textit{ Expert in science and science
	education, guiding with knowledge and pedagogy.} 
    \end{description}

\item \textbf{Art \& Music} --- Topics that focus on artistic expression,
    music generation, and creative arts, including tools for creating,
    editing, and experiencing art and music.

    \begin{description}
    \item[\textbf{Example:}] \textit{ The best AI artwork generator for
	effortlessly creating highly detailed illustrations or photos
	without requiring any design expertise. This designer GPT will
	produce stunning images with just a simple prompt.}
    \end{description}

\item \textbf{Repair \& Maintenance} --- Topics that cover tools and guides
    for repairing, maintaining, and troubleshooting various equipment or
    systems, ensuring optimal functionality.

    \begin{description}
    \item[\textbf{Example:}] \textit{ The GPT trains building owners about
	the operation and maintenance of Powergate VFDs.} 

    \end{description}

\item \textbf{Agriculture \& Farming} --- Topics that involve topics
    related to agriculture, farming practices, and agri-tech innovations,
    focusing on sustainable and efficient food production.

    \begin{description}
    \item[\textbf{Example:}] \textit{ A knowledgeable agriculture mentor,
	providing practical farming advice.}
    \end{description}

\item \textbf{Culture \& History} --- Topics that encompass cultural
    knowledge, historical information, and heritage preservation, promoting
    awareness and appreciation of diverse cultures and histories. 

    \begin{description}
    \item[\textbf{Example:}] \textit{ Build history from a divergence
	point. As you build history, ask it to tell you stories.}
    \end{description}

\item \textbf{Legal \& Documentation} --- Topics that pertain to legal
    assistance, documentation, compliance, and regulatory information,
    offering guidance on legal matters and document management.

    \begin{description}
    \item[\textbf{Example:}] \textit{ Expert legal assistant for drafting
	motions, proposed orders, and other legal documents.} 
    \end{description}

\end{itemize}

\begin{figure}
    \centering
    \includegraphics[width=1\linewidth]{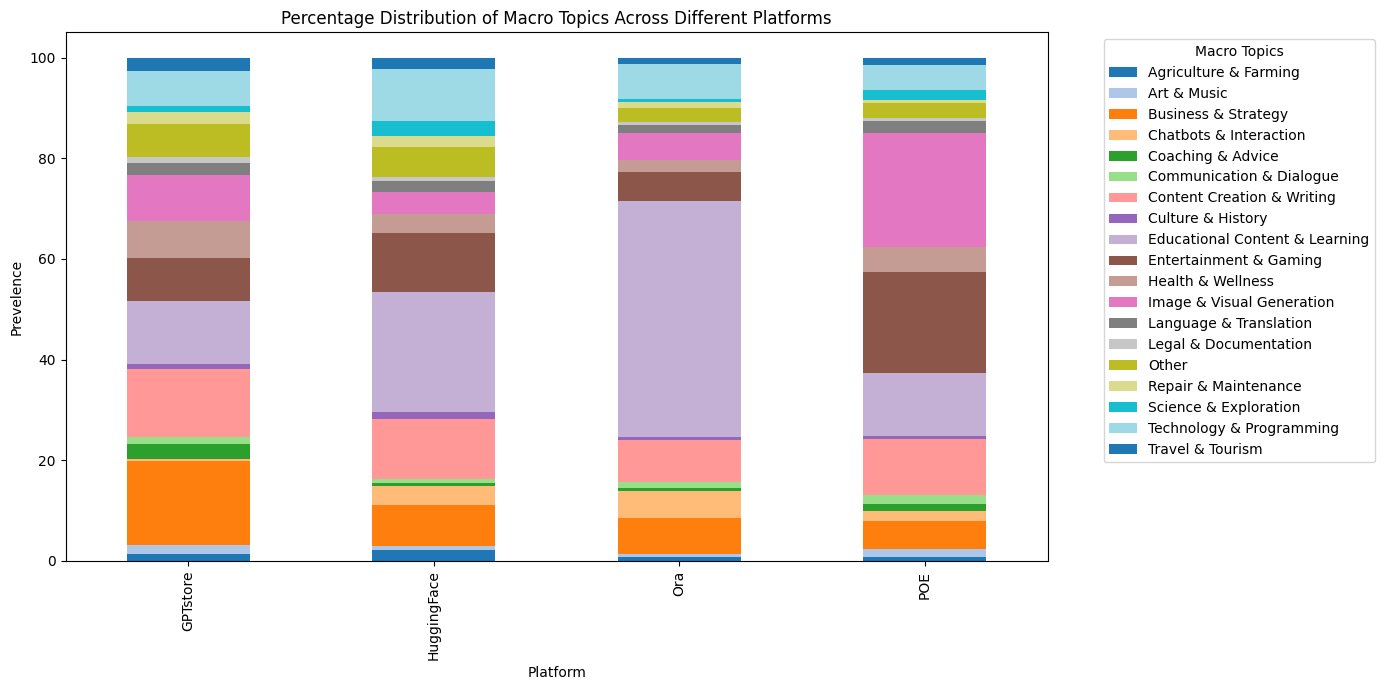}
    \caption{Distribution of Macro-topics across different platforms}
    \label{fig:enter-label}
\end{figure}

Fig.~\ref{fig:enter-label} shows the differences in the distribution of macro-topics across the various platforms.
We can see that ``Educational Content \& Learning'' and ``Content Creation \& Writing'' consistently rank among the top categories across all surveyed platforms. This underscores a strong interest by Promptware developers in tools for education and content generation within the Large Language Model (LLM) ecosystem.

GPTstore appears to focus significantly on professional and business needs, with ``Business \& Strategy'' being its most prevalent category. This is closely followed by topics related to content creation and education, indicating a user base that leverages LLMs for business development and the enhancement of professional skills.

HuggingFace, known for its open-source contributions, emphasizes education. The data reflects a community that values learning and creativity, with a significant engagement in entertainment and technology programming, highlighting its diverse user interests.

Ora stands out with nearly half of its offerings centered around educational tools, suggesting it serves as a hub for academic and educational applications. This focus points to its use in formal learning environments or self-learning scenarios, catering to a wide range of learners.

POE distinguishes itself with a focus on visual generation and gaming, indicating a community centered around graphic design, gaming, and interactive entertainment. The creative domain seems to be a major attraction for users on this platform.

Conversely, categories such as ``Chatbots \& Interaction'', ``Culture \& History'', and ``Legal \& Documentation'' appear less common than others across all platforms. While our findings reveal the presence of legal-related Promptware, further work is needed to determine whether this reflects emerging regulation trends.
For instance, ``Chatbots \& Interaction'' could gain prominence due to the rapid advancement of multimodal conversational agents and the growing demand for personalized, task-specific virtual assistants across industries such as customer service, education, and mental health. Similarly, the demand for ``Legal \& Documentation'' tools may rise as organizations explore AI for legal research, compliance automation, and contract analysis --- especially as regulatory interest in responsible AI and explainability grows. The broader push toward domain-specific applications of LLMs suggests that topics that are less prominent today may become more central as both technical capabilities and real-world deployment scenarios continue to evolve.

Additionally, the variance in category prevalence highlights each platform's unique market positioning and strategic priorities. GPTstore's business-centric approach contrasts with HuggingFace's focus on education and open-source contributions, Ora's role as an educational resource hub, and POE's creative and entertainment orientation. These differences reflect the platforms' attempts to cater to distinct user segments and the diverse ways in which LLMs are being leveraged to meet specific needs and interests.

Overall, the findings illustrate a dynamic and varied LLM ecosystem, with each platform serving different aspects of user demand, ranging from professional development and strategic business applications to educational tools, creative arts, and entertainment. This diversity underscores the versatility of LLMs in addressing a wide array of use cases and points to evolving trends that may shape the future of FMware development and adoption.

\subsection{RQ2 Results}
\label{sec:rq2}

\textbf{RQ2: What technical challenges do developers encounter when building and maintaining FMware?}  

To answer this question, we applied a combined approach of label-based and AI-assisted classification to categorize issues from the FMware repositories. This process produced a distribution of 146,570 issues across 15 distinct categories from GitHub, as presented in Table~\ref{tab:classification_distribution}. The classification reveals several trends in the types of challenges and development focuses within the FMware ecosystem.

\begin{table}[ht]
\centering
\caption{Classification Distribution}
\begin{tabular}{lrr}
\toprule
\textbf{Classification} & \textbf{Count} & \textbf{Percentage (\%)} \\
\midrule
Bug Reports                          & 42,661 & 30.69 \\
Core Functionality                   & 30,364 & 21.85 \\
Miscellaneous                        & 11,632 & 8.37 \\
Documentation                        & 10,949 & 7.88 \\
Support \& Questions                 & 8,219  & 5.91 \\
Feature Requests                     & 6,706  & 4.82 \\
Infrastructure \& Deployment         & 6,584  & 4.74 \\
Project Management                   & 4,410  & 3.17 \\
User Experience (UX)                 & 3,794  & 2.73 \\
Performance \& Optimization          & 3,774  & 2.72 \\
API \& Integration                   & 3,395  & 2.44 \\
Quality Assurance \& Testing         & 2,850  & 2.05 \\
Platform Compatibility               & 1,861  & 1.34 \\
Internationalization \& Localization & 902    & 0.65 \\
Security                             & 887    & 0.64 \\
\bottomrule
\end{tabular}
\label{tab:classification_distribution}
\end{table}

The most frequently occurring categories are ``Bug Reports'' (30.69\%) and ``Core Functionality'' (21.85\%), which together account for over half of the collected issues. These categories primarily reflect ongoing efforts to address technical challenges related to FMware reliability and system behavior. While this distribution may seem unsurprising --- bug fixing and feature development are typically among the most common issue types in many software systems (e.g.,~\cite{herbold2022problems}) --- their dominance reinforces that FMware development shares many baseline maintenance concerns with traditional software.

What distinguishes FMware, however, becomes clearer upon deeper analysis. In the next phase, we examine these two categories using topic modeling to uncover the specific types of challenges developers face. This reveals a more nuanced picture of FMware-specific concerns such as memory allocation failures, model loading errors, and collaborative review bottlenecks.

Other notable categories include ``Support \& Questions'' (5.91\%), reflecting a substantial need for community-driven troubleshooting and knowledge sharing, and ``Feature Requests'' (4.82\%), which suggest continued interest in expanding FMware capabilities. These categories highlight the collaborative and evolving nature of FMware development, where users actively seek help and propose enhancements to adapt the tools to their needs. 

Overall, the distribution of issue classifications provides a nuanced view of the challenges that developers encounter in FMware development. ``Bug Resolution'', ``Core Functionality'', and ``Documentation'' stand out as primary areas requiring ongoing attention, while categories like ``User Experience'' and ``Infrastructure'' suggest a developing interest in FMware systems.

To validate the generalization of our methodology, we further analyzed LLM-related issues on GitHub to construct the taxonomy.
Unlike prior work such as that of Chen et al.~\cite{chenxiang}, which directly sampled and categorized issues from official GitHub repositories of major LLM vendors --- specifically, Meta AI, Google, and OpenAI --- our study first applied a combined approach of label-based and AI-assisted classification to categorize a broader range of FMware-related issues. This process produced a distribution of 146,570 issues across 15 distinct categories, as shown in Table~\ref{tab:classification_distribution}. From this classification, we identified ``Bug Reports'' (30.69\%) and ``Core Functionality'' (21.85\%) as the two most prevalent categories, together accounting for over 50\% of all issues. 
To gain deeper insights, we selected these top two categories and applied topic modeling, resulting in 26,650 posts for ``Bug Reports'' and 24,484 posts for ``Core Functionality''. 
While Chen et al.~\cite{chenxiang} constructed a taxonomy directly from sampled issues, our approach ensures a structured categorization before topic modeling, providing a more systematic and scalable analysis of the FMware ecosystem. 
The taxonomy developed in our study captures a diverse range of software engineering challenges specific to FMware, complementing prior research while extending the understanding of LLM-related development issues.

\subsubsection{Bug Reports Topic Modeling Results}
\label{sec:results:bug_reports}

In this section, we present the results from topic modeling on the ``Bug
Reports'' dataset. We identified 30 distinct topics, each highlighting a
specific type of issue frequently encountered by developers.

\begin{figure}[htbp]
  \centering
  \includegraphics[width=1\columnwidth]{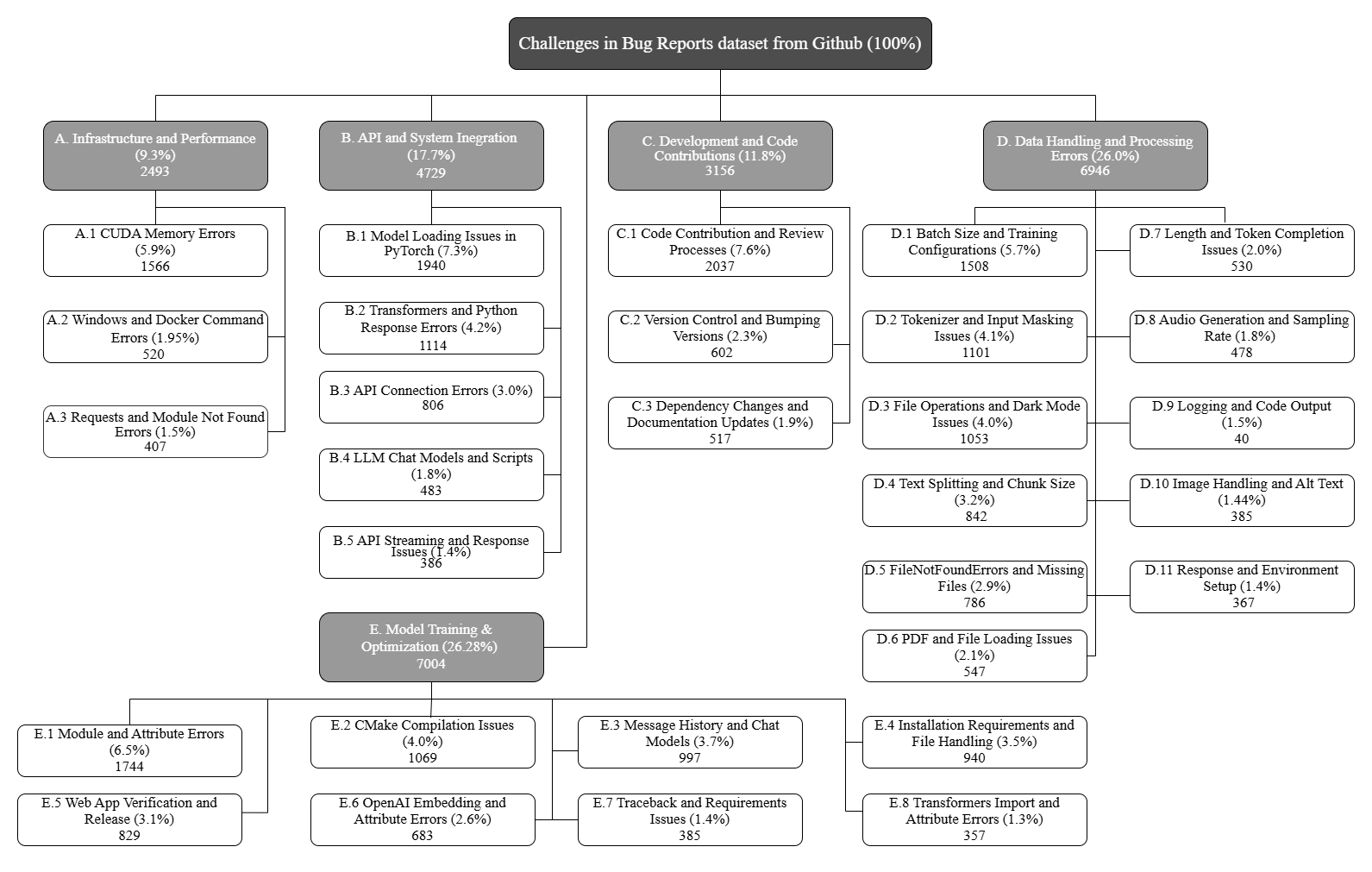}
  \caption{Challenge Taxonomy for Bug Reports Dataset Built from GitHub Issues}
  \label{fig:bug_reports_distribution}
\end{figure}

To systematically analyze the recurring issues in FMware development, we
categorized the identified topics from the bug reports dataset into a
structured taxonomy as \textbf{Figure~\ref{fig:bug_reports_distribution}}.
This taxonomy classifies bug-related challenges into five major categories:
``Infrastructure \& Hardware Issues'', ``API \& Model Execution'',
``Development \& Code Contribution Errors'', ``Data Handling \& Processing
Errors'', and ``Module \& Installation Issues''. Each category encompasses
distinct technical issues that developers frequently encounter, such as
``CUDA Memory Errors and Hardware Compatibility Issues to API Connection
Failures'', ``Tokenization Problems'', ``Dependency Conflicts'', and
``Module Import Errors''.


\subsubsection{Distribution of Challenge Topics in Bug Reports Dataset}

In this section, we present the distribution of challenge topics identified in the bug reports dataset. Our analysis covered a wide range of issues, with a total of 30 distinct topics. Figure~\ref{fig:bug_reports_distribution} provides an overview of the distribution of these topics.

To highlight key findings, we group the most prevalent topics under their corresponding high-level categories and summarize their significance below:
\begin{itemize}

\item
\textbf{General Errors} are dominated by \textbf{CUDA Memory Errors (A.1, 5.9\%)}, indicating persistent struggles with GPU memory constraints. These errors occur frequently in large-scale deep learning workflows, where memory optimization is crucial for model training and inference. Developers often encounter ``out of memory'' failures, limiting their ability to efficiently run high-performance models.

\item 
\textbf{API and Model Execution} is primarily impacted by \textbf{Model Loading Issues in PyTorch (B.1, 7.3\%)}, reflecting common challenges in restoring models from checkpoints and resolving format incompatibilities. The frequency of these errors suggests that ensuring smooth model deployment and loading remains a critical concern, particularly when integrating large FMware-based models into production environments.

\item
\textbf{Development and Code Contributions} see the most reports in \textbf{Code Contribution and Review Processes (C.1, 7.6\%)}, emphasizing the importance of collaborative software engineering. Given that many FMware repositories rely on contributions from a broad developer community, maintaining high-quality code through structured reviews, pull request management, and version control is essential.

\item 
\textbf{Data Handling and Processing} is largely shaped by \textbf{Batch Size and Training Configurations (D.1, 5.7\%)}, highlighting difficulties in tuning hyperparameters for model training. Developers frequently face challenges in optimizing batch sizes to balance computational efficiency and GPU memory usage, underscoring the technical complexity of training deep learning models in constrained environments.

\item
\textbf{Module and Installation Issues} are most represented by \textbf{Module and Attribute Errors (E.1, 6.5\%)}, which stem from dependency conflicts, incorrect environment configurations, and missing module imports. These errors often disrupt development workflows, requiring significant debugging efforts to resolve compatibility issues between software components.

\end{itemize}

The prominence of these topics highlights critical areas of focus within
FMware development, where improvements in memory management, collaborative
practices, and training setup could enhance efficiency and scalability.
Other notable topics include ``Model Loading Issues in PyTorch'', ``Module
and Attribute Errors'', and ``Transformer Response Errors'', each
representing key technical challenges encountered by developers.

This distribution provides insights into the key technical and operational
challenges encountered by developers in the FMware ecosystem. By focusing
on the most prevalent issues, such as ``Memory Management'', ``Code
Contribution Processes'', and ``Training Configuration'', this study offers
a foundation for targeted improvements in FMware development tools and
workflows.

\subsubsection{Core Functionality Topic Modeling Results}
\label{sec:results:core_functionality}

In this section, we present the results from performing topic modeling on the Core Functionality dataset. We identified 30 distinct topics, each highlighting a specific type of issue frequently encountered by developers when working on core functionalities of FMware systems.

To better understand the challenges faced by developers when working with the core functionalities of FMware systems, we categorized the identified topics into a structured taxonomy, presented in \textbf{Figure~\ref{fig:core_distribution}}. This taxonomy follows the same high-level categories as those used in the bug reports dataset (\textbf{Section~\ref{sec:results:bug_reports}}), ensuring consistency and enabling direct comparisons between different types of challenges encountered in FMware development. By maintaining a uniform taxonomy, we can analyze whether certain challenges are more prevalent in bug reports (indicating frequent failures and debugging needs) or in core functionality discussions (reflecting broader architectural and optimization concerns).

To highlight key findings, we group the most prevalent topics under their corresponding high-level categories and summarize their significance below:

\begin{figure}[htbp]
    \centering
    \includegraphics[width=1\columnwidth]{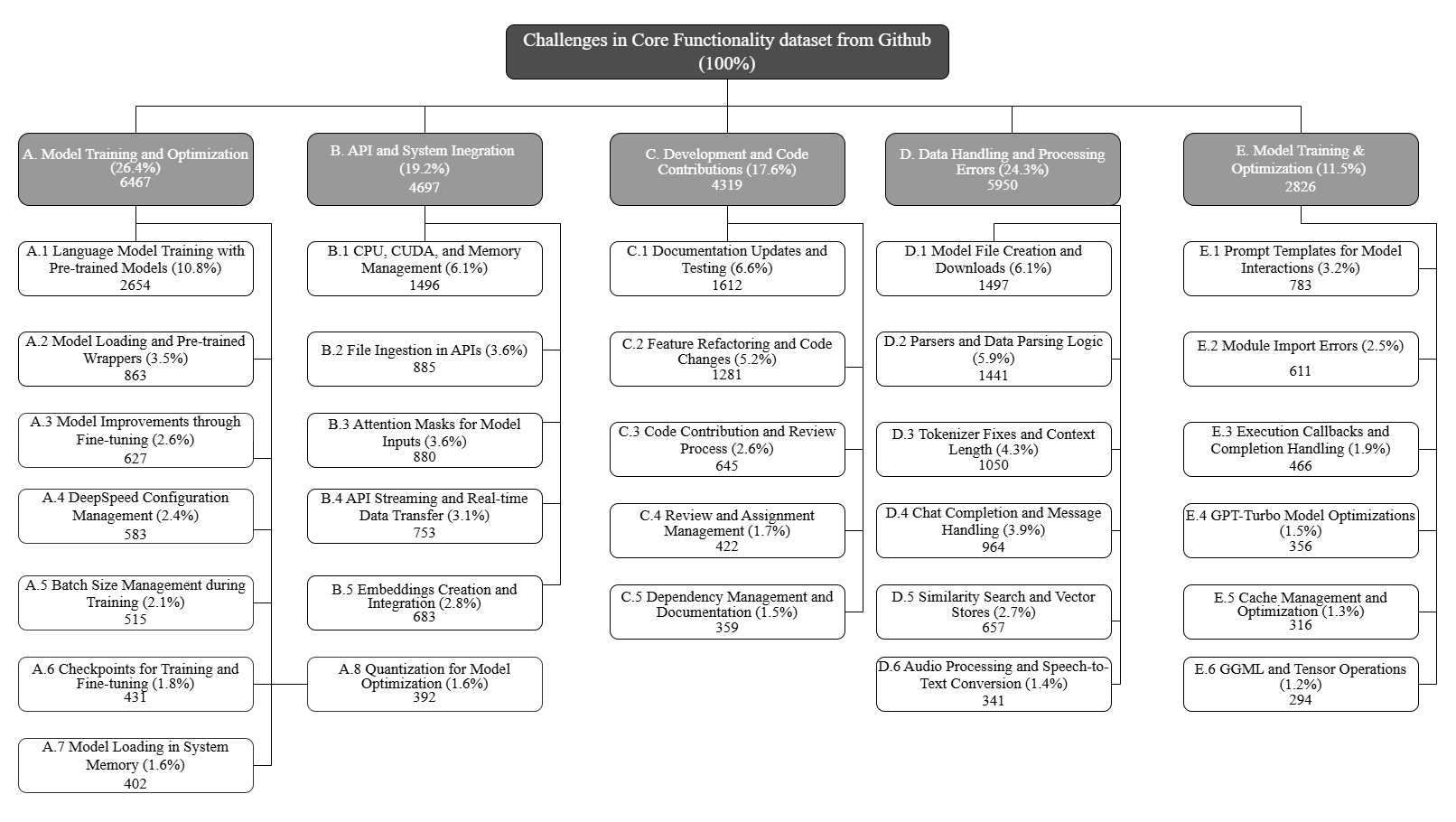}
    \caption{Challenge Taxonomy for Core Functionalities Dataset Built from GitHub Issues}
    \label{fig:core_distribution}
\end{figure}

\begin{itemize}
\item 
\textbf{Model Training \& Optimization} is dominated by \textbf{Language Model Training with Pre-trained Models (A.1, 10.8\%)}, indicating a strong focus on model fine-tuning, checkpoint management, and optimization techniques. Developers frequently encounter challenges in customizing pre-trained models for specific applications, ensuring training efficiency, and balancing resource constraints.

\item
\textbf{API \& System Integration} is primarily impacted by \textbf{CPU, CUDA, and Memory Management (B.1, 6.1\%)}, reflecting concerns around efficient execution of models within constrained hardware environments. This topic includes managing memory allocation, handling CUDA-related errors, and optimizing system performance when running FMware models at scale.

\item
\textbf{Development \& Code Contribution} sees the most reports in \textbf{Documentation Updates and Testing (C.1, 6.6\%)}, highlighting the need for accurate documentation and rigorous testing in open-source FMware projects. Developers often focus on refining documentation to keep up with evolving features and ensuring that new contributions undergo proper validation.

\item
\textbf{Data Processing \& Handling} is largely shaped by \textbf{Model File Creation and Downloads (D.1, 6.1\%)}, emphasizing the complexities of managing pre-trained models, ensuring compatibility, and handling large model files efficiently. Given the rapid expansion of FMware-based applications, smooth handling of model assets is crucial for reducing overhead and streamlining workflows.

\item
\textbf{Infrastructure \& Performance} is most represented by \textbf{Prompt Templates for Model Interactions (E.1, 3.2\%)}, which underscores the growing importance of structured prompts in enhancing model usability and response consistency. Optimizing prompt templates ensures that language models generate reliable, meaningful, and context-aware responses across different applications.

\end{itemize}

The prominence of these topics highlights key areas of focus within FMware
development, where improvements in ``Model Training'', ``System
Performance'', ``Documentation Practices'', and ``Data Handling Workflows''
can enhance efficiency and scalability. Other notable topics include
``Feature Refactoring and Code Changes'', ``API Streaming and Real-Time
Data Transfer'', and ``Execution Callbacks and Completion Handling'', each
addressing essential aspects of FMware functionality.

This structured distribution provides insights into the core technical and operational challenges encountered by FMware developers. The structured breakdown provides a comprehensive view of the prevalent challenges developers face in FMware core functionality.

\subsection{RQ3 Results}
\label{sec:rq3}

\textbf{RQ3: Which types of FMware issues require the most time to resolve?}  

To address this question, we analyze the time required to resolve issues across different datasets to identify which topics demand greater developer effort. Our study includes two primary datasets from GitHub repositories, as used in RQ2: the \textbf{Bug Reports} dataset, consisting of 26,650 issues, and the \textbf{Core Functionality Issues} dataset, consisting of 30,355 issues. For each issue, we calculate the solving time as the difference between the opening and closing timestamps, measured in hours. We then compute the median solving time per topic to reduce the influence of outliers and better reflect typical resolution durations. This approach provides a robust comparison of topic-level complexity and developer burden across both datasets.  

\begin{figure}
    \centering
    \includegraphics[width=1\linewidth]{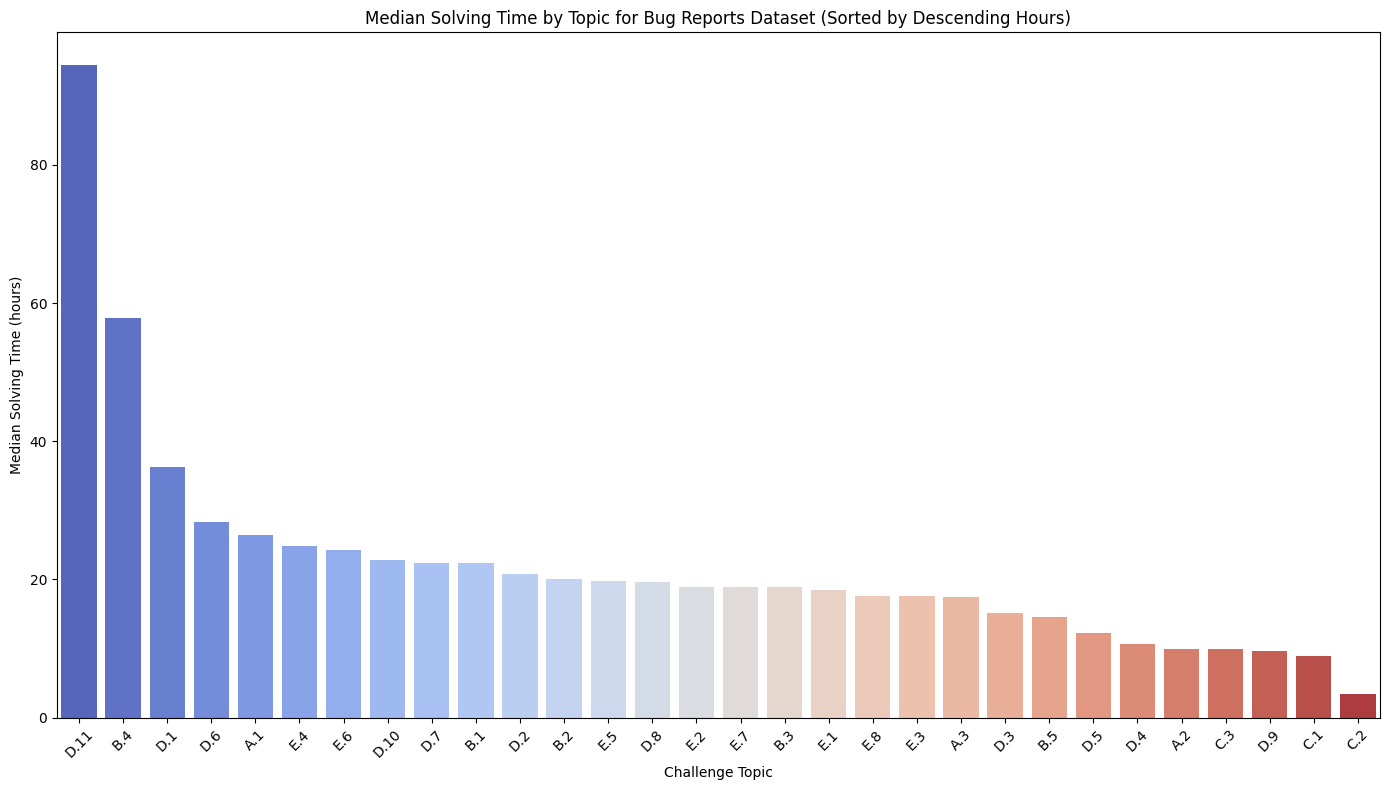}
    \caption{Median Solving Time by Topic for Bug Reports Dataset}
    \label{fig:bug_time}
\end{figure}

\subsubsection{Results for Bug Report Solving Time}
To visualize the time demands of different bug types, Figure \ref{fig:bug_time} presents the median resolution time (in hours) for each bug report topic, sorted from highest to lowest. This helps identify which categories of FMware bugs typically require more developer effort and time to resolve.
Based on median solving times, three topics stand out as the most time-consuming to resolve in the Bug Reports dataset: \textbf{Response and Environment Setup (D.11, 1.4\%, 367 issues)}, \textbf{LLM Chat Models and Scripts (B.4, 1.8\%, 483 issues)}, and \textbf{Batch Size and Training Configurations (D.1, 5.7\%, 1508 issues)}. These topics reflect complex and time-intensive development challenges in FMware.

\textbf{Response and Environment Setup (D.11)} has the highest median solving time. This topic includes issues related to configuring and troubleshooting environments for CUDA and PyTorch, such as version mismatches or missing components. The elevated resolution time highlights the intricacies involved in correctly setting up dependencies and runtime environments, which are often critical for FMware deployment.

\textbf{LLM Chat Models and Scripts (B.4)} ranks second in median resolution time. It includes tasks related to script orchestration, module integration, and managing chat-based LLM interactions. The longer solving time underscores the complexity of coordinating various components in conversational AI applications and debugging across dependent modules.

\textbf{Batch Size and Training Configurations (D.1)} comes third, with challenges involving tuning hyperparameters such as batch sizes for efficient model training. These tasks frequently require empirical tuning and careful consideration of hardware constraints, especially in large-scale training setups.

We also analyze the mean solving time for this dataset. The top three topics by mean are \textbf{LLM Chat Models and Scripts (B.4, 1.8\%, 483 issues)}, \textbf{OpenAI Embedding and Attribute Errors (E.6, 2.6\%, 683 issues)}, and \textbf{PDF and File Loading Issues (D.6, 2.1\%, 547 issues)}. While B.4 is ranked highly in both metrics, D.11 drops significantly in the mean-based ranking --- indicating the presence of extreme outliers in other categories that pull up their averages. This highlights why median-based analysis can provide a more robust view of central developer effort, especially when distributions are skewed.
Together, these findings indicate that resolving infrastructure setup, conversational model integration, and training configuration issues requires extended effort. This highlights the need for better tooling, clearer documentation, and automation for environment and training setup processes in FMware development.

\subsubsection{Results for Core Function Dataset Solving Time}

For the Core Functionality dataset, the top three topics with the highest median solving times are: \textbf{Language Model Training with Pre-trained Models (A.1, 10.8\%, 2654 issues)}, \textbf{Model Loading and Pre-trained Wrappers (A.2, 3.5\%, 863 issues)}, and \textbf{Quantization for Model Optimization (A.8, 1.6\%, 392 issues)}. These topics primarily fall under model training and optimization efforts.

\textbf{Language Model Training with Pre-trained Models (A.1)} has the highest median resolution time. This reflects the intensive and often iterative process of training and fine-tuning models. Issues under this topic include adjusting training loops, managing datasets, and ensuring convergence, which are all critical for model performance.

\textbf{Model Loading and Pre-trained Wrappers (A.2)} follows closely. This topic captures issues related to restoring pre-trained checkpoints and integrating wrapper interfaces. Long solving times here suggest that compatibility and version mismatches in model loading remain significant pain points in FMware development.

\textbf{Quantization for Model Optimization (A.8}) ranks third in time intensity. Developers face challenges optimizing models for efficient deployment --- often through quantization --- while ensuring that accuracy is not degraded. This process requires both domain-specific expertise and extensive experimentation.

We also analyze the mean solving time for the Core Function dataset. The top three topics by mean are \textbf{Code Contribution and Review Processes (C.3, 2.6\%, 645 issues)}, \textbf{Similarity Search and Vector Stores (D.5, 2.7\%, 657 issues)}, and \textbf{Prompt Templates for Model Interactions (E.1, 3.2\%, 783 issues)}. This shift from A-group topics to C, D, and E groups in mean-based rankings suggests that while training-related issues are consistently time-consuming on average, developer collaboration, vector index tuning, and prompt engineering occasionally require disproportionately long resolution times --- likely due to coordination delays or complex iterative debugging.

While these findings highlight which types of FMware issues are most time-consuming to resolve, our analysis is limited to FMware projects. A natural next step would be to compare these patterns against traditional software repositories to determine whether the observed delays are unique to FMware or reflect broader trends in software engineering practice. At the same time, our results show that model-related engineering tasks dominate the most time-consuming FMware issues in core functionality. This indicates a strong need for improved tools, automation frameworks, and development guidelines to support training, loading, and optimizing large foundation models.

\begin{figure}
    \centering
    \includegraphics[width=1\linewidth]{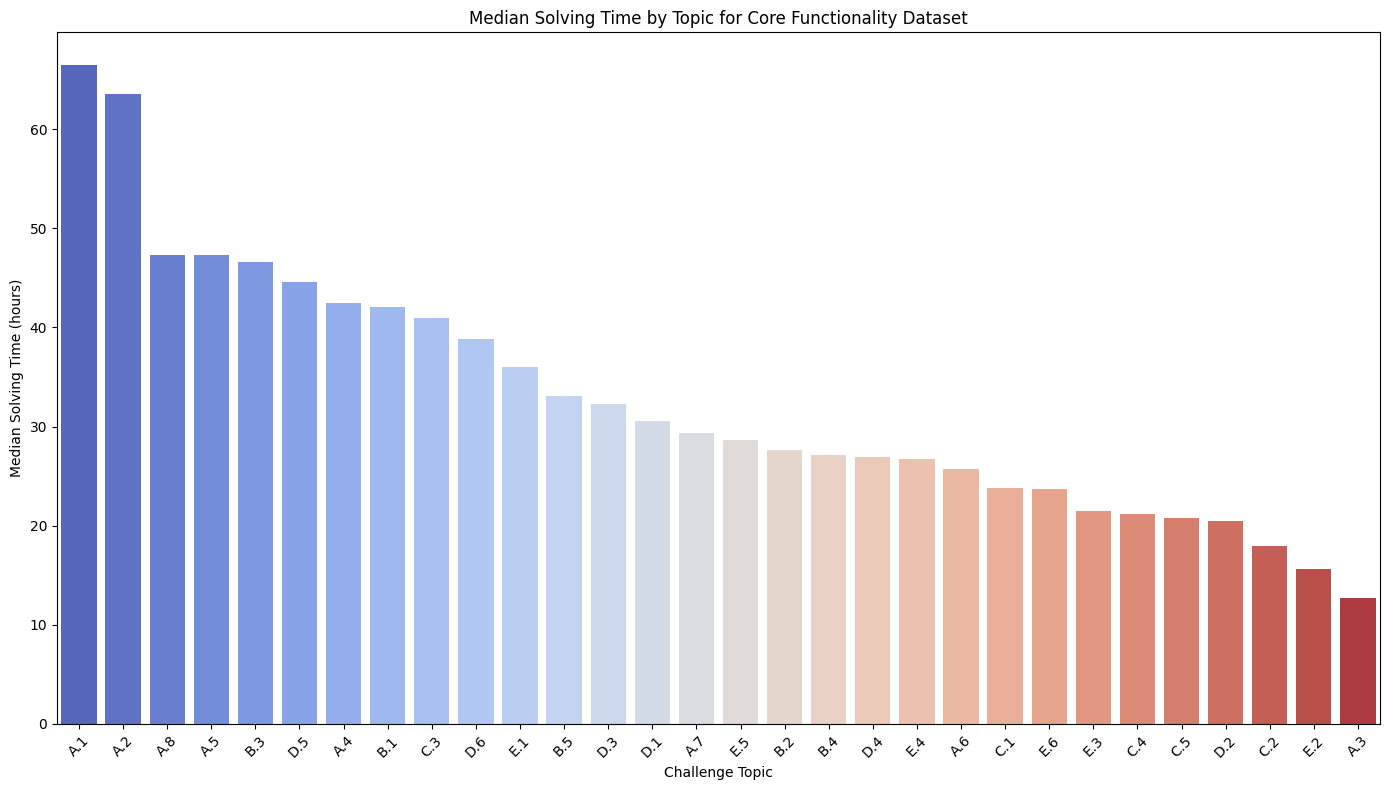}
    \caption{Median Solving Time by Topic for Core Functionality Dataset}
    \label{fig:core_time}
\end{figure}
\section{Discussion and Implications}
\label{sec:implications}
Our research offers insights for a wide range of FMware practitioners, including researchers, platform developers, and tool developers. By exploring the prevalent macro-topics across different FMware platforms, we highlight key areas of focus within the ecosystem and identify potential gaps in FMware development.

\subsection{Implications for Researchers}
\label{sec:implications:researchers}
\vspace{-0.5em} 

\paragraph{Cross-Platform Research Opportunities:}

One of the most critical insights from our study is the variation in macro-topic distribution across different FMware platforms, such as HuggingFace, GPTstore, Ora, and POE. This indicates that each platform serves distinct user bases and needs, ranging from business strategy to creative arts. However, the cross-platform implications of these differences remain underexplored. Researchers could focus on investigating how developers and users interact across multiple FMware platforms and whether their experiences differ based on the platform they use.

Moreover, cross-platform tool interoperability remains a largely untapped area. As developers and enterprises increasingly adopt multiple FMware solutions, understanding how tools and models can be seamlessly integrated across platforms will be key. For example, research into creating frameworks that allow for the portability of FMware models and applications from cloud-based solutions (e.g., HuggingFace) to on-premise environments (e.g., GitHub-hosted FMware) could reduce the friction developers face when migrating or deploying across different infrastructures.

Given the prevalence of topics related to Educational Content \& Learning and Content Creation \& Writing, there is also a need for researchers to focus on improving the usability and collaboration within FMware platforms. A key challenge across platforms is ensuring that FMware tools are intuitive and accessible for users at different skill levels. Researchers could explore user experience (UX) design for FMware interfaces, particularly how to make complex FMware functionalities more accessible to non-experts, such as educators or content creators.

Additionally, collaborative FMware development is an area ripe for exploration. As FMware applications become more complex and involve larger teams, understanding how developers collaborate on FMware projects, share models, and manage dependencies across platforms could lead to new tools and practices that improve team-based development in the FMware ecosystem.

\paragraph{Benchmarking and Evaluation Frameworks for FMware:}
Finally, our study identifies a gap in the benchmarking and evaluation of FMware tools. While existing FMware platforms cater to diverse use cases, there is no standardized framework for evaluating the effectiveness or performance of FMware models across different platforms. Researchers could work on developing evaluation benchmarks that assess FMware performance across multiple dimensions, such as efficiency, scalability, and user satisfaction. These benchmarks would help researchers and practitioners alike to better understand which FMware platforms and tools are most effective for particular tasks and could guide future development efforts.

\subsection{Validating Challenges in Production-Grade FMware Engineering}

Our empirical findings provide strong validation for the curated set of engineering challenges outlined in Rethinking Software Engineering in the Foundation Model Era by Hassan et al. \cite{rethinking}. Their work presents a conceptual taxonomy of barriers that hinder the development and deployment of production-ready FMware, including issues related to prompt instability, environment fragility, integration complexity, and team coordination. Our study substantiates many of these challenges through large-scale analysis of GitHub issues, while also revealing gaps between theory and current practice.

One of the most prominent challenges identified in the prior work involves managing prompts and achieving reliable execution behavior. Our results support this concern through the emergence of the topic “Prompt Templates for Model Interactions” (E.1) in the Core Functionality dataset, which comprises 3.2\% of all issues in that set. Developers frequently struggle with prompt modularity, output consistency, and parameter sensitivity --- factors that contribute to unpredictable behavior in LLM-based applications, as supported by prior evaluations of explainable fault localization and prompt behavior analysis in LLM-based debugging workflows.\cite{dis1,dis3,dis4,dis2} The substantial solving time associated with this topic further reflects the trial-and-error nature of prompt engineering, thus confirming the practical significance of prompt-related instability. This aligns with findings from Kang et al.~\cite{kang2025explainable}, who demonstrate how explainable, LLM-driven debugging can help interpret and improve prompt outcomes in fault-prone development scenarios.

Another critical area emphasized in the prior literature is the fragility of runtime environments and infrastructure. Our analysis highlights this as a major obstacle: the topic “Response and Environment Setup” (D.11) from the Bug Reports dataset and “CPU, CUDA, and Memory Management” (B.1) from the Core Functionality dataset together capture frequent reports of dependency mismatches, CUDA compatibility issues, and resource bottlenecks. These topics had high median solving times and appeared consistently across diverse repositories, underscoring the difficulty developers face when configuring FMware environments for training or inference. This finding echoes recent debugging studies that emphasize runtime brittleness and the need for interpretable debugging mechanisms for LLM-based systems.\cite{dis3,dis4} 

We also validate the challenge of model integration and adaptation, particularly for those attempting to load and fine-tune pre-trained FMs. The most prevalent topic in the Core Functionality dataset --- “Language Model Training with Pre-trained Models” (A.1, 10.8\%) --- emerged as a clear bottleneck. Issues in this category include checkpoint loading failures, incompatibilities across framework versions, and iterative tuning failures, echoing similar challenges reported by Chen et al.\cite{chen2025llmtraining} in their empirical study of LLM training systems. Similar topics such as “Model Loading and Pre-trained Wrappers” (A.2) and “Quantization for Model Optimization” (A.8) further demonstrate the complexity involved in customizing or extending FMs for new applications. These results empirically support the claim that productionizing FMs entails substantial engineering investment far beyond API access or prompt design.

Another key challenge addressed in the “Rethinking” paper is collaborative development. In our Bug Reports dataset, “Code Contribution and Review Process” (C.1) was the single most frequent issue category (7.6\%). This indicates the significant effort required for managing pull requests, conducting reviews, and maintaining code quality in FMware projects, many of which are developed by distributed teams. Although prior discussions often emphasize model-centric challenges, our findings suggest that process-centric concerns --- especially those related to team coordination and contribution workflows --- are equally critical in practice.

While our findings reinforce many of the challenges highlighted in prior work, certain areas appear underrepresented in the GitHub issue data. Notably, we found limited evidence of runtime hallucination, closed-loop evaluation, or reliability-related concerns in our topic models. These omissions may stem from the nature of GitHub as a platform for developer-centric discussions; user-facing issues related to factual correctness or usability may instead surface in forums, product feedback channels, or post-deployment logs. Similarly, topics such as “Security and Compliance” and “Internationalization” appeared infrequently in both datasets, despite being listed as high-priority concerns for enterprise-grade FMware systems. This discrepancy suggests that while these topics are conceptually important, they may not yet be central to the daily workflows of open-source FMware developers.

Beyond validating existing challenges, our findings highlight a tooling gap for diagnosing and debugging FMware systems. While GitHub issues reflect frustration with environment setup, loading checkpoints, and parameter tuning, recent research has proposed LLM-based debuggers that emulate human reasoning~\cite{dis3}, test-free fault localization techniques~\cite{dis2}, and explainable fault analysis pipelines~\cite{dis1}, all of which could be adapted to improve observability and reliability in FMware development.

In summary, our empirical analysis validates many of the engineering challenges proposed by Hassan et al., including those related to prompt brittleness, infrastructure setup, model adaptation, and collaborative workflows. At the same time, our results reveal a gap between conceptual frameworks and the practical realities reflected in issue tracking systems. These findings provide actionable guidance for future research: while current tooling should focus on improving model integration and environment robustness, further studies are needed to capture runtime behaviors, trustworthiness, and cross-platform evaluation practices that may not yet be visible through GitHub alone.

\section{Threats to Validity}
\label{sec:threat}

In this section, we discuss potential threats to the validity of our study and how they could impact our findings.

\textbf{Internal Validity:} Several factors may introduce biases in our topic modeling and classification approach. First, \textbf{topic modeling limitations} may arise due to the use of BERTopic, which is sensitive to pre-processing choices, hyperparameter selection, and the underlying embedding model. As a result, the extracted topics might not always reflect the true concerns of FMware developers. Second, \textbf{ the accuracy of the data labeling} could be affected by AI-assisted classification of issues, particularly for unlabeled GitHub issues, leading to possible misclassification errors that affect the accuracy of the topic distribution. Third, \textbf{preprocessing loss of information} may have occurred, as the removal of certain stop words such as “error” or “bug” could have unintentionally stripped meaningful context from the discussion of the issue.

\textbf{External Validity:} The extent to which our findings generalize beyond the analyzed dataset is another important consideration. \textbf{Generalizability across the FMware ecosystem} is a potential concern, as our data set consists primarily of GitHub-hosted FMware projects, which may not fully represent the challenges faced in cloud-based platforms such as HuggingFace, Poe, or Ora. Differences in infrastructure, deployment environments, and user interactions may lead to distinct issue distributions across platforms. Moreover, GitHub-hosted FMware may not reflect industrial FMware systems that are closed-source and subject to different organizational constraints. Community-specific biases also exist, as our dataset primarily captures developer-reported concerns on GitHub, potentially underrepresenting user-centric issues (e.g., usability and accessibility challenges) that are discussed more frequently on community forums.

\textbf{Construct Validity:} The consistency of our taxonomy across different datasets presents a potential threat to construct validity. While we maintain the same upper categories for both the Bug Reports dataset and the Core Functionality dataset to ensure comparability, taxonomy consistency across datasets may overlook domain-specific differences in issue types. Certain challenges unique to core functionality discussions may not align perfectly with those in bug reports, potentially leading to misclassification or loss of granularity. In addition, our measure of issue resolution time may be affected by factors beyond technical complexity, such as project activity levels, contributor availability, or project management practices. These factors could bias the interpretation of which FMware issues are genuinely more difficult to resolve.

\textbf{Reliability:} The reproducibility of our findings is subject to several factors. \textbf{Reproducibility of results} is influenced by BERTopic hyperparameter settings, AI-assisted classification methods, and dataset selection. Different runs with varying parameters or data sources may yield slight variations in topic distributions. Additionally, \textbf{changes in the FMware landscape over time} present a challenge, as the dataset reflects a snapshot of current FMware development practices. As FMware tools and frameworks evolve, new challenges may emerge that are not captured in this study, requiring ongoing investigation to track shifting trends in FMware development.


\section{Summary}
\label{sec:conclusion}

This study presents the first large-scale empirical analysis of FMware development --- the processes of building and maintaining software systems around foundation models --- by examining both cloud-based Promptware platforms and on-premise GitHub-hosted repositories. 
We addressed three core focus areas: identifying the domains in which FMware is most commonly applied, characterizing the key technical challenges developers encounter, and determining which types of issues demand the most time and effort to resolve.

Our analysis shows that FMware platforms are concentrated in education, content creation, and business applications, reflecting the broad applicability of Foundation Models across diverse domains. 
From a development perspective, we identified persistent technical challenges in memory management, dependency handling, tokenizer configuration, and collaborative workflows. 
Issues involving code contribution, similarity search, and prompt engineering required the most time to resolve, underscoring the complexity of building robust and scalable FMware systems.

By analyzing more than 25,000 FMware applications and 140,000 GitHub issues, this study provides actionable insights for researchers, practitioners, and tool builders. 
The findings highlight the need for better tooling, improved debugging support, and more effective workflows to sustain the rapidly growing FMware ecosystem. 
As FMware development continues to accelerate, addressing these challenges will be essential to unlocking the full potential of Foundation Models in software engineering.
\section{Conflict of Interest}
\label{sec:conflict}
\vspace{-0.5em}

The authors declare that they have no conflict of interest.
\section{Data Availability Statement}
\label{sec:availability}

The datasets generated and analyzed during this study are available in the replication package~\cite{wang2025fmware}.
\section{Declarations}

\subsection{Funding}
This work was supported by the \textbf{NSERC Discovery Grant program, Canada} (Award No.~RGPIN-04183-2018), awarded to Michael W. Godfrey.

\subsection{Ethical Approval}
Not applicable.

\subsection{Informed Consent}
Not applicable.

\subsection{Author Contributions}
All authors contributed to the study conception and design. Data collection and analysis were performed by Zitao Wang, with feedback and validation from Zhimin Zhao and Michael W. Godfrey. The first draft of the manuscript was written by Zitao Wang, and all authors commented on previous versions of the manuscript. All authors read and approved the final manuscript.

\subsection{Data Availability Statement}
The datasets and scripts generated and analyzed during the current study are available in the replication package\cite{wang2025fmware}.

\subsection{Conflict of Interest}
The authors declare that they have no conflict of interest.

\subsection{Clinical Trial Number}
Clinical trial number: not applicable.

\bibliography{ref}
\bibliographystyle{spmpsci}

\appendix
\section{Appendix}

\section*{Stop Words Used in RQ1: PromptWare Descriptions}
\label{app:stopwordsrq1}
For RQ1, we removed general English nouns and vague domain terms that did not help differentiate topics, alongside frequent but uninformative platform-specific words. Below is the final stop word list used after manual review and coherence tuning.

\paragraph{General Content Words Removed:}
\begin{quote}
\small
ability, abilities, accident, accidents, acknowledgement, action, actions, activities, activity, advantage, advantages, alternative, alternatives, announcement, announcements, anomaly, anomalies, answer, answers, appreciation, approach, approaches, article, articles, assistance, author, behavior, behaviour, benefit, benefits, bit, bits, body, case, cases, cause, causes, challenge, challenges, change, changes, check, choice, choices, collection, com, combination, concern, concerns, confirmation, confusion, consideration, context, couple, couples, course, courses, cross, day, days, demand, desire, detail, details, difference, differences, difficulties, difficulty, disadvantage, disadvantages, discrepancies, discrepancy, discussion, discussions, dislike, distinction, effect, end, enquiries, enquiry, evidence, example, examples, exception, exceptions, existence, exit, expectation, experience, expert, experts, explanation, explanations, fact, facts, favorite, favorites, feedback, feedbacks, forum, forums, future, goal, goals, guarantee, guidance, guideline, guide, guides, guy, guys, harm, hello, help, hour, hours, idea, ideas, individual, individuals, info, information, inquiries, inquiry, insight, intelligence, intent, interest, introduction, investigation, invitation, issues, kind, kinds, lack, learning, level, levels, look, looks, lot, lots, luck, major, manner, manners, manual, mark, means, meaning, method, methods, minute, minutes, month, months, need, needs, number, numbers, offer, one, ones, opinion, opinions, outcome, part, parts, past, people, person, persons, perspective, perspectives, place, places, point, points, post, posts, practice, practices, problem, problems, product, products, proposal, proposals, purpose, purposes, py, qa, questions, reason, reasons, result, results, scenario, scenarios, science, second, seconds, section, sense, shortcoming, shortcomings, show, shows, situation, software, solution, solutions, start, study, stuff, success, summary, summaries, surprise, support, supports, task, tasks, technique, techniques, technologies, technology, term, terms, tip, tips, thank, thanks, thing, things, thought, thoughts, three, title, today, tomorrow, total, trouble, troubles, truth, try, tutorial, tutorials, two, understand, understanding, usage, use, user, users, uses, view, viewpoint, way, ways, weakness, weaknesses, week, weeks, word, words, work, workaround, workarounds, works, yeah, year, years, yesterday
\end{quote}

\paragraph{Platform-Specific and Domain-Redundant Words Removed:}
\begin{quote}
\small
GPT, gpt, gpts, store, gptstore, chatgpt, bot, bots, com, p, post, a, openai, open, ai, AI, https, http, assistant, assistants, knowledge, custom, think, star, want, try, trying, year, kid, chat, chats
\end{quote}

These words were identified through term frequency analysis and manually reviewed to exclude overly generic or FMware-branded terms that reduced topic distinctiveness.

\section*{Stop Words Used in RQ2: GitHub Issues}
\label{app:Rq2stopwords}

For RQ2, we removed both generic English terms and GitHub- or FMware-specific words that frequently appeared in issue discussions but did not meaningfully contribute to topic separation. This included discussion-related filler words, platform references, and terms overly tied to specific bug descriptions (e.g., ``bug'', ``fix'', ``error'').

\paragraph{Platform-Specific and Redundant Discussion Terms:}
\begin{quote}
\small
GPT, gpt, gpts, store, gptstore, chatgpt, bot, bots, gpt4, gpt-4, com, p, post, a, openai, open, ai, AI, https, http, assistant, assistants, knowledge, custom, think, star, want, try, trying, year, kid, chat, chats, issue, issue number, issues, issues\_number, issue\_number, image, images, video, creator, guidance, friend, wingman, photo, invest, investment, game, play, fitness, therapy, seo, learn, writing, essay, search, url, urls, index, error, fix, fixe, fixed, bug, bugs, buggy
\end{quote}

\paragraph{General Content Words Removed:}
\begin{quote}
\small
ability, accident, acknowledgement, action, activities, advantage, article, assistance, author, behavior, benefit, bit, cause, challenge, check, choice, com, detail, discrepancy, discussion, dislike, effect, example, exception, experience, expert, fact, feedback, goal, guideline, idea, info, inquiry, method, minute, need, number, offer, part, point, problem, reason, result, second, solution, support, task, technique, term, thing, tutorial, usage, way, week, work
\end{quote}

These words were selected through manual review and topic coherence evaluation. Diagnostic keywords such as ``error'' and ``bug'' were removed only after verifying their overrepresentation in nearly all issue types, limiting their value in distinguishing topics.

\end{document}